\documentstyle[aps,preprint,tighten]{revtex}

\begin{document}
\draft

\title{$\Lambda$(1405) as a multiquark state}
\author{Seungho Choe\thanks{E-mail: schoe@physics.adelaide.edu.au}}
\address{Special Research Centre for the Subatomic Structure of Matter\\
University of Adelaide, Adelaide, SA 5005, Australia}
\maketitle
\begin{abstract}

In the QCD sum rule approach
we predict the $\Lambda$ (1405) mass by choosing
the  $\pi^0\Sigma^0$ multiquark interpolating field.
It is found that the mass is about 1.419 GeV from $\Pi_1 (q^2)$ sum rule
which is more reliable than $\Pi_q (q^2)$ sum rule,
where $\Pi_q (q^2)$ and $\Pi_1 (q^2)$ are two invariant functions
of the correlator $\Pi (q^2)$. We also present the sum rules
for the $K^+ p$ and the $\pi^+\Sigma^+$ multiquark states,
and compare to those for the $\pi^0\Sigma^0$ multiquark state.
The mass of the $\Lambda$ (1600) can be also reproduced
in our approach.

\end{abstract}
\vspace{1cm}

\section{Introduction}

There have been many works to study properties of the $\Lambda$(1405)
and its roles in nuclear physics
\cite{1,2,3,4,5,6,7,8,9,10,11,12,13,14,15,16}.
However, its nature is not revealed completely, i.e., an ordinary
three quark state or a $\bar{K}N$ bound state or
a mixed state of the previous two possibilities.
For a historical review and another references, see \cite{prd96}.
The $\Lambda$ (1405) has long been considered as a $\bar{K}N$ bound state
\cite{sakurai60,dwr67} since it is just below the $\bar{K}N$ threshold.
Thus, it is assumed to be a candidate
of hadronic molecules, which are weakly-bound
states of two or more hadrons
\footnote{For a review of hadronic molecules, see \cite{barnes94}}.
In this paper we use the QCD sum
rule approach to predict a mass of the $\Lambda$(1405)
considering a multiquark picture.

QCD sum rule\cite{svz79,rry85,narison89}
is one of powerful tools to extract the hadron
properties directly from QCD.
Applications of this approach to the $\Lambda$ (1405)
were done in Ref. \cite{liu84,leinweber90,kl96}, respectively.
In Ref.\cite{liu84}
it  was shown that the five quark operators, which
correspond to a baryon and pseudoscalar meson or a baryon and vector
meson, have great contribution to the mass of the $\Lambda$(1405).
This result is consistent with an analysis from the
MIT bag model\cite{strottman79}.
On the other hand, Leinweber\cite{leinweber90}
 obtained a good fit to the splitting between
the $\Lambda$ (1405) and the $\Lambda$ (1520) using a three-quark
interpolating field.
Recently, as an another approach,
 Kim and Lee \cite{kl96} proposed a three-quark interpolating field
with a covariant derivative for the $\Lambda$ (1405)
according to the quark configuration of that in the bag model or
the non-relativistic quark model.

On the basis of Liu's result\cite{liu84}
one can  assume the $\Lambda$(1405) as
 a $\bar{K}N$ hadronic molecular state (five-quark state)
and investigate the possibility following the same procedures
in Ref.\cite{choe97}.
However, in the case of the $K^- p$
multiquark state there are no exchange diagrams at tree level
such as those
shown in Ref.\cite{choe97}. Thus, at tree level it is impossible
to determine
whether there is a binding effect or not in the $K^- p$
multiquark state.
On the other hand the $\Lambda$ (1405)
is observed in the mass spectrum of the $\pi\Sigma$ channel (I=0),
so it would be interesting to
get a mass assuming the $\pi\Sigma$ multiquark state
instead of the $\bar{K}N$
multiquark state. Among three $\pi\Sigma$ channels,
only the $\pi^0\Sigma^0$
multiquark state has exchange diagrams
in contrast to the $\pi^+\Sigma^-$ and $\pi^-\Sigma^+$
multiquark states.

 In Sec. \ref{multi1}, we present a QCD sum rule prediction of
the mass of the $\Lambda (1405)$ taking into account
the  $\pi^0\Sigma^0$ interpolating field, and
in Sec. \ref{multi2} we
compare the results of the
$\pi^0\Sigma^0$ multiquark state
with those of the $K^+ p$ multiquark state (I=1)
and the $\pi^+\Sigma^+$ multiquark state (I=2), respectively.
We discuss uncertainties in our sum rules
and summarize our results in Sec. \ref{discuss}.

\section{QCD sum rule for $\pi^0\Sigma^0$ multiquark}{\label{multi1}}

Let's consider the following correlator:
\begin{eqnarray}
\Pi (q^2) = i \int d^4x e^{iqx}\langle T ( J(x) \bar{J}(0) )\rangle ,
\end{eqnarray}
where $J = \epsilon_{abc}(\bar{u}_e i\gamma^5 u_e
			     -\bar{d}_e i\gamma^5 d_e)
([u_a^T C\gamma_\mu s_b]\gamma^5\gamma^\mu d_c
 + [d_a^T C\gamma_\mu s_b]\gamma^5\gamma^\mu u_c) $
corresponds
 to the interpolating field for the  $\pi^0\Sigma^0$ multiquark state.
$u$, $d$ and $s$ are the up, down and strange quark fields, and
$a,b,c,e$ are color indices.
 $T$ denotes the transpose in Dirac space, and $C$ is the charge
conjugation matrix.
The OPE side has two structures:
\begin{eqnarray}
\Pi^{OPE} (q^2) =\Pi_{q}^{OPE} (q^2) {\bf \rlap{/}{q}}
		+\Pi_{1}^{OPE} (q^2) {\bf 1} ,
\end{eqnarray}
where
\begin{eqnarray}
\Pi_{q}^{OPE} (q^2) = &-& \frac{11}{\pi^8 ~2^{17} ~3^2 ~5^2 ~7} q^{10} ln(-q^2)
   +  \frac{11~m_s^2}{\pi^8 ~2^{17} ~3^2 ~5} q^8 ln(-q^2)
\nonumber\\
&-& \frac{11 ~m_s}{\pi^6 ~2^{13} ~3^2 ~5}
\langle\bar{s}s\rangle  q^6 ln (-q^2)
- \frac{3}{\pi^4 ~2^{11}}
\langle\bar{q}q\rangle^2 q^4 ln(-q^2)
\nonumber\\
&+& \frac{3 ~m_s^2}{\pi^4 ~2^{8}}
\langle\bar{q}q\rangle^2 q^2 ln(-q^2)
- \frac{3 ~m_s}{\pi^2 ~2^{7}}
\langle\bar{q}q\rangle^2 \langle\bar{s}s\rangle ln(-q^2)
\nonumber\\
&-& \frac{11}{2^3 ~3^3}
\langle\bar{q}q\rangle^4 \frac{1}{q^2} ,
\label{ope1_q}
\end{eqnarray}
and
\begin{eqnarray}
\Pi_{1}^{OPE} (q^2) =  &-& \frac{11 ~m_s}{\pi^8 ~2^{18} ~3^2 ~5^2}
	q^{10} ln(-q^2)
+ \frac{11}{\pi^6 ~2^{15} ~3^2 ~5}
\langle\bar{s}s\rangle q^8 ln(-q^2)
\nonumber\\
&+& \frac{11 ~m_s^2}{\pi^6 ~2^{14} ~3^2}
\langle\bar{s}s\rangle q^6 ln(-q^2)
- \frac{49 ~m_s}{\pi^4 ~2^{9} ~3^2}
\langle\bar{q}q\rangle^2 q^4 ln (-q^2)
\nonumber\\
&+& \frac{3}{\pi^2 ~2^6}
\langle\bar{q}q\rangle^2 \langle\bar{s}s\rangle q^2 ln(-q^2)
- \frac{m_s^2}{\pi^2 ~2^6 ~3}
(14\langle\bar{q}q\rangle^3
- 9\langle\bar{q}q\rangle^2\langle\bar{s}s\rangle) ln(-q^2)
\nonumber\\
&-& \frac{m_s}{2^4 ~3^3}
(44\langle\bar{q}q\rangle^4
 + 3 \langle\bar{q}q\rangle^3\langle\bar{s}s\rangle) \frac{1}{q^2} .
\label{ope1_1}
\end{eqnarray}
Here, we let
$m_u$ = $m_d$ = 0 $\neq$ $m_s$ and
$\langle\bar{u}u\rangle$ = $\langle\bar{d}d\rangle$
$\equiv$ $\langle\bar{q}q\rangle$ $\neq$ $\langle\bar{s}s\rangle$.
We neglect the contribution of gluon condensates and
concentrate on tree diagrams,
and assume the vacuum saturation hypothesis
to calculate quark condensates of higher dimensions.
Similar calculation was done
in Kodama ${\it et ~al.}$'s H-dibaryon sum rules\cite{koh94}.
Note that in $\Pi_q^{OPE}$
we neglect the term which is proportional to
 $m_s^2 \langle\bar{q}q\rangle^4$ ${1\over q^4}$
in order to keep the same order of power corrections
as in $\Pi_1^{OPE}$,
but its contribution is less than 1 \%.

The OPE sides have the following form:
\begin{eqnarray}
\Pi^{OPE}_{q,1} (q^2) &=& a ~q^{10} ln(-q^2) + b ~q^8 ln(-q^2)
+ c ~q^6 ln(-q^2)
+ d ~q^4 ln(-q^2)
\nonumber \\
&+& e ~q^2 ln(-q^2)+ f ~ln(-q^2) + g ~\frac{1}{q^2} ,
\end{eqnarray}
where $a, b, c, \cdots, g$ are constants. Then, we parameterize the
phenomenological sides as
\begin{eqnarray}
\frac{1}{\pi} Im \Pi^{Phen}_{q} (s) &=& \lambda^2 \delta(s-m^2) +
	     [-a~s^5 - b~s^4 - c~s^3 - d~s^2 - e~s - f] \theta(s~-~s_0) ,
\nonumber\\
\frac{1}{\pi} Im \Pi^{Phen}_{1} (s) &=& \lambda^2 m \delta(s-m^2) +
	     [-a~s^5 - b~s^4 - c~s^3 - d~s^2 - e~s - f] \theta(s~-~s_0) ,
\end{eqnarray}
where $m$ is the mass of the $\pi^0\Sigma^0$ multiquark state
and $s_0$ a continuum threshold for each sum rules.
$\lambda$ is the coupling strength of the interpolating field
to the physical $\Lambda$ (1405) state.
After Borel transformation we obtain the mass of the $\pi^0\Sigma^0$
multiquark state
from $\Pi_q$ and $\Pi_1$, respectively.
The mass $m$ is given by
\begin{eqnarray}
m^2 &=& M^2 \times
\nonumber\\
&\{&-720a (1-\Sigma_6)
- \frac{120b}{M^2} (1-\Sigma_5)
- \frac{24c}{M^4} (1-\Sigma_4)
\nonumber\\
&-& \frac{6d}{M^6} (1-\Sigma_3)
- \frac{2e}{M^8} (1-\Sigma_2)
- \frac{f}{M^{10}} (1-\Sigma_1) \}
~/
\nonumber\\
&\{&-120a (1-\Sigma_5)
- \frac{24b}{M^2} (1-\Sigma_4)
- \frac{6c}{M^4} (1-\Sigma_3)
\nonumber\\
&-& \frac{2d}{M^6} (1-\Sigma_2)
-\frac{e}{M^8} (1-\Sigma_1)
- \frac{f}{M^{10}} (1-\Sigma_0)
- \frac{g}{M^{12}} \} ,
\label{mass}
\end{eqnarray}
where
\begin{eqnarray}
\Sigma_i = \sum_{k=0}^i \frac{s_0^k}{k~! ~(M^2)^k} ~e^{-\frac{s_0}{M^2}}  .
\end{eqnarray}
In Eq. (\ref{mass}) the continuum contribution is large, so this formula
has large uncertainties.
We can not find a plateau for
the mass of the $\pi^0\Sigma^0$
multiquark state in the relevant Borel region.
Fig. \ref{fig_ps0s} shows the masses from $\Pi_q$ and $\Pi_1$ sum rules,
where the continuum threshold is taken to be $s_0$ = 2.789 GeV$^2$
considering
the next $\Lambda$ (1670) particle.
The solid line is the mass prediction from $\Pi_q$ sum rule
while the dotted line is that from $\Pi_1$ sum rule.
It seems that $\Pi_1$ sum rule
is more stable than $\Pi_q$ sum rule.
This is consistent
with a recent work by Jin and Tang \cite{jt97}.
They showed that the chiral-odd sum rule ($\Pi_1$ sum rule)
is generally
more reliable than the chiral-even sum rule ($\Pi_q$ sum rule)
because the positive- and negative-parity excited-state
contributions partially cancel each other in $\Pi_1$ sum rule, but add up
in $\Pi_q$ sum rule.
In Fig. \ref{fig_ps0s}
there is a plateau for large Borel mass, but it is a trivial result
from our crude model on the phenomenological side.

Figs. \ref{fig_ps0v} (a), (b)
denote the dependencies of the predicted mass
on the s-quark mass and
the s-quark condensate. Only the results from
$\Pi_1$ sum rule are shown in the limited Borel region at the
same continuum threshold, i.e., $s_0$ = 2.789 GeV$^2$.
It seems that the SU(3) symmetry breaking effect
is small in our sum rules. On the other hand  the mass is rather
dependent on the quark condensate as shown
in Fig. \ref{fig_ps0v} (c).

First, before getting the mass of the $\Lambda$ (1405),
we can apply the same procedures as in
Ref. \cite{choe97} and determine whether there is a binding effect
in the $\pi^0\Sigma^0$ multiquark state.
In the followings we introduce an effective threshold $s_0$
which will be used to obtain the mass of the $\Lambda$ (1405).
The OPE sides can be rewritten as
\begin{eqnarray}
\Pi^{OPE}_{q,1} (q^2) = N_c^2 ( {\rm ~2 ~loop-type}
			 + {1 \over N_c} \times {\rm ~1 ~loop-type} ) ,
\end{eqnarray}
where 2 loop-type means the contribution from diagrams of
Fig. \ref{fig_loop} (a),
and 1 loop-type the contribution from diagrams of
Fig. \ref{fig_loop} (b).
$N_c$ is the number of color, and in
Eqs. (\ref{ope1_q}), (\ref{ope1_1}) above
we take $N_c$ = 3.
Note that in Figs. \ref{fig_loop} (a), (b)
we present only some typical diagrams.
Then, we can use the same strategy in Ref.\cite{choe97}:
First, consider 2 loop-type only and vary the continuum threshold $s_0$ and
the Borel interval $M^2$
in order that the mass should be 1.328 GeV
(the sum of a pion and a $\Sigma$ mass).
The Borel interval $M^2$ is restricted by
the following conditions as usual:
The lower limit of $M^2$ is determined as the value at which the power
correction is below than 30\%. The upper limit is determined
as the value where the continuum contribution in the mass prediction is
less than 50\%.
Second, consider all diagrams (1 loop-type + 2 loop-type) and
get a new mass $m^\prime$ at the same $s_0$ and Borel interval
$M^2$ which are obtained
from the first step.
Third, compare	$m^\prime$ with 1.328 GeV. If $m^\prime$ is less than
1.328 GeV, it can be one signature for a molecular-like multiquark state.

Figs. \ref{fig_ps0} (a), (b) show the dependence of the mass 
on the Borel mass $M^2$
in both sum rules, i.e., $\Pi_q$ and $\Pi_1$ sum rule, assuming
$\langle\bar{q}q\rangle$ = --(0.230 GeV)$^3$,
$\langle\bar{s}s\rangle$ = 0.8 ~$\langle\bar{q}q\rangle$,
and $m_s$ = 0.150 GeV.
The solid line is the mass prediction including
2 loop-type diagrams only ($m$),
and the dotted line is the mass with all diagrams ($m^\prime$).
Because there is no plateau in the valid Borel
region for both cases, we determine
the mass as an average value in that Borel interval.
The results are given in Table ~\ref{table_qc}.
It shows that there is a binding effect in 
$\Pi_q$ sum rule while a repulsive effect in $\Pi_1$ sum rule.
The binding effect is about 32 MeV in $\Pi_q$ sum rule.
However, in $\Pi_1$ sum rule the average mass is
slightly larger than the $\pi\Sigma$ threshold.
The difference between the average mass and the threshold value is
about 3 MeV as shown in the table.

We also present the dependence of $m^\prime$
on different quark condensates in Table ~\ref{table_qc}.
The mass is rarely changed for different quark condensates.
The dependencies of $m^\prime$ on the s-quark mass and the
s-quark condensate are given in Tables ~\ref{table_sqm}
and ~\ref{table_sc}, respectively.
In our approach, if we take another s-quark mass and/or s-quark
condensate, then the continuum threshold and/or the Borel
interval also should be changed to reproduce the mass $m$ (1.328 GeV).
Thus, the variation of the mass $m^\prime$ is small.

As we said, $\Pi_q$ sum rule
is less reliable than $\Pi_1$ sum rule, 
and if we take a large value of the quark condensate
 the average mass becomes slightly
smaller than the threshold in the case of $\Pi_1$ sum rule .
Hence, at present stage it is difficult to determine
whether there is the binding effect in the $\pi^0\Sigma^0$ multiquark state.

Now, let us determine the mass of the $\Lambda$ (1405).
Fig. \ref{fig_coup} shows the coupling strength $\lambda^2$
from $\Pi_q$ and $\Pi_1$ sum rule at the threshold
$s_0$ = 3.122 GeV$^2$ and $s_0$ = 3.012 GeV$^2$, respectively.
There is the maximum point in the case of $\Pi_1$ sum rule while not in
the case of $\Pi_q$ sum rule.
Then, we can determine the mass of the $\Lambda$ (1405)
when the coupling strength becomes the maximum value
at the threshold $s_0$ obtained in the previous calculation.
In Table \ref{mass_qc} the calculated masses are presented for
different quark condensates.
In Tables \ref{mass_sqm} and \ref{mass_sc} the dependencies of the
mass on the s-quark mass and the s-quark condensate
are given.
The predicted mass 1.419 GeV for
$\langle\bar{q}q\rangle$ = --(0.230 GeV)$^3$
is  similar to the experimental value.

\section{QCD sum rule for $K^+ p$ and $\pi^+\Sigma^+$
multiquark}{\label{multi2}}

In this section
we apply the previous approach	to the
$K^+ p$ multiquark state (I=1) and
the $\pi^+\Sigma^+$ multiquark state (I=2)
each other.
These channels are not exist as a resonance contrary to the
$\pi^0\Sigma^0$ channel. Thus, the results for these multiquark
states should be different from those for the $\pi^0\Sigma^0$
multiquark state. In the case of the $K^+ p$ multiquark state
the OPE sides are given as follows.
\begin{eqnarray}
\Pi_{q}^{OPE} (q^2) = &-& \frac{1}{\pi^8 ~2^{16} ~3^2 ~5 ~7} q^{10} ln(-q^2)
   +  \frac{m_s^2}{\pi^8 ~2^{16} ~3^2} q^8 ln(-q^2)
\nonumber\\
&+& \frac{m_s}{\pi^6 ~2^{12} ~3^2 ~5}
(7\langle\bar{q}q\rangle - 5\langle\bar{s}s\rangle ) q^6 ln (-q^2)
\nonumber\\
&-& \frac{1}{\pi^4 ~2^{10} ~3^2}
(2\langle\bar{q}q\rangle^2
+ 7\langle\bar{q}q\rangle \langle\bar{s}s\rangle) q^4 ln(-q^2)
\nonumber\\
&+& \frac{m_s^2}{\pi^4 ~2^{8} ~3^2}
(4\langle\bar{q}q\rangle^2
- 7\langle\bar{q}q\rangle\langle\bar{s}s\rangle) q^2 ln(-q^2)
\nonumber\\
&+& \frac{m_s}{\pi^2 ~2^{5} ~3^2}
(5\langle\bar{q}q\rangle^3
- \langle\bar{q}q\rangle^2 \langle\bar{s}s\rangle) ln(-q^2)
\nonumber\\
&-& \frac{5}{2^2 ~3^3}
\langle\bar{q}q\rangle^3 \langle\bar{s}s\rangle  \frac{1}{q^2} ,
\label{ope2_q}
\end{eqnarray}
\begin{eqnarray}
\Pi_{1}^{OPE} (q^2) = &+& \frac{1}{\pi^6 ~2^{14} ~3^2}
\langle\bar{q}q\rangle q^8 ln(-q^2)
	- \frac{5m_s^2}{\pi^6 ~2^{12} ~3^2}
	\langle\bar{q}q\rangle q^6 ln(-q^2)
\nonumber\\
&-& \frac{m_s}{\pi^4 ~2^9 ~3^2}
(4\langle\bar{q}q\rangle^2
- 5\langle\bar{q}q\rangle \langle\bar{s}s\rangle ) q^4 ln (-q^2)
\nonumber\\
&+& \frac{1}{\pi^2 ~2^5 ~3^2}
(7\langle\bar{q}q\rangle^3
+ 2\langle\bar{q}q\rangle^2 \langle\bar{s}s\rangle) q^2 ln(-q^2)
\nonumber\\
&-& \frac{m_s^2}{\pi^2 ~2^4 ~3^2}
(7\langle\bar{q}q\rangle^3
- \langle\bar{q}q\rangle^2\langle\bar{s}s\rangle) ln(-q^2)
\nonumber\\
&-& \frac{m_s}{2^3 ~3^3}
(20\langle\bar{q}q\rangle^4
- 7\langle\bar{q}q\rangle^3\langle\bar{s}s\rangle) \frac{1}{q^2} ,
\label{ope2_1}
\end{eqnarray}
and  for the $\pi^+\Sigma^+$ multiquark state we get
\begin{eqnarray}
\Pi_{q}^{OPE} (q^2) = &-& \frac{1}{\pi^8 ~2^{16} ~3^2 ~5 ~7} q^{10} ln(-q^2)
   +  \frac{m_s^2}{\pi^8 ~2^{16} ~3^2} q^8 ln(-q^2)
\nonumber\\
&-& \frac{m_s}{\pi^6 ~2^{12} ~3^2}
\langle\bar{s}s\rangle	q^6 ln (-q^2)
- \frac{1}{\pi^4 ~2^{10}}
\langle\bar{q}q\rangle^2 q^4 ln(-q^2)
\nonumber\\
&+& \frac{m_s^2}{\pi^4 ~2^{7}}
\langle\bar{q}q\rangle^2 q^2 ln(-q^2)
- \frac{m_s}{\pi^2 ~2^{6}}
\langle\bar{q}q\rangle^2 \langle\bar{s}s\rangle ln(-q^2)
\nonumber\\
&-& \frac{5}{2^2 ~3^3} \langle\bar{q}q\rangle^4 \frac{1}{q^2} ,
\label{ope3_q}
\end{eqnarray}
and
\begin{eqnarray}
\Pi_{1}^{OPE} (q^2) = &-& \frac{m_s}{\pi^8 ~2^{17} ~3^2 ~5}
	q^{10} ln(-q^2)
     + \frac{1}{\pi^6 ~2^{14} ~3^2} \langle\bar{s}s\rangle q^8 ln(-q^2)
\nonumber\\
&+& \frac{5~m_s^2}{\pi^6 ~2^{13} ~3^2} \langle\bar{s}s\rangle  q^6 ln (-q^2)
- \frac{m_s}{\pi^4 ~2^8}
\langle\bar{q}q\rangle^2  q^4 ln(-q^2)
\nonumber\\
&+& \frac{1}{\pi^2 ~2^5}
\langle\bar{q}q\rangle^2\langle\bar{s}s\rangle q^2 ln(-q^2)
+ \frac{m_s^2}{\pi^2 ~2^5}
\langle\bar{q}q\rangle^2 \langle\bar{s}s\rangle ln(-q^2)
\nonumber\\
&-& \frac{5 ~m_s}{2 ~3^3} \langle\bar{q}q\rangle^4 \frac{1}{q^2}  ,
\label{ope3_1}
\end{eqnarray}
where we take $N_c$ = 3 as in the previous  $\pi^0\Sigma^0$
multiquark sum rules (hereafter  $\pi^0\Sigma^0$  sum rules).

Comparing Eqs. (\ref{ope3_q}) and (\ref{ope3_1}) to
 Eqs. (\ref{ope2_q}) and (\ref{ope2_1}) each other one can easily find that
the formulas
are exactly the same in the SU(3) symmetric limit.
Then, two SU(3) symmetry breaking parameters
($m_s$ and $\gamma$ $\equiv$
$\langle\bar{s}s\rangle \over \langle\bar{q}q\rangle$ -- 1 )
give different characteristics
between the $K^+ p$ and the $\pi^+\Sigma^+$ sum rules.
Fig. \ref{fig_kp} and Fig. \ref{fig_ps} are the mass predictions,
where we use $\langle\bar{q}q\rangle$ = --(0.230 GeV)$^3$
,  $\langle\bar{s}s\rangle$ = 0.8 ~$\langle\bar{q}q\rangle$,
$m_s$ = 0.150 GeV. We let
$m$ = $m_{K^+}$ + $m_p$ = 1.435 GeV and
$m$ = $m_{\pi^+}$ + $m_{\Sigma^+}$ = 1.329 GeV
for the $K^+ p$ sum rules and the $\pi^+ \Sigma^+$ sum rules, respectively.
They show the same pattern for each sum rules.
In Tables~\ref{table_kp} and \ref{table_ps} we present the average masses
in the valid Borel interval.
It shows that there is a binding effect in $\Pi_q$ sum rule for both
multiquark states. The binding effect is about 60 MeV and  100 MeV
for the $K^+ p$ and the $\pi^+\Sigma^+$ multiquark state, respectively.
On the other hand,
 in $\Pi_1$ sum rule there is a repulsive effect
for both states although the magnitude is very small in our approach,
and this is the same as the experimental result \cite{dw82}.
Thus, also in these two cases $\Pi_1$ sum rule seems more reliable
than $\Pi_q$ sum rule.

We can apply the previous method to get the mass
of the $K^- p$ multiquark state. The $K^- p$ sum rule
is the same as the $K^+ p$ sum rule when only 2-loop diagrams
are considered.
Using the same threshold in Table \ref{table_kp}, i.e.,
$s_0$ = 3.852 GeV$^2$
we get 1.589 GeV from $\Pi_1$ sum rule.
We also obtain the mass of the $\pi^-\Sigma^+$ multiquark state
from the $\pi^+\Sigma^+$ sum rule.
The predicted mass at the same
$s_0$ = 3.852 GeV$^2$
is 1.606 GeV. In the case of the $\pi^0\Sigma^0$ sum rule we get 1.625 GeV
at this threshold.
It is interesting to note that these values are very similar to each other,
and close to that of the $\Lambda$ (1600) which can decay to both $\bar{K}N$
and $\pi\Sigma$ channels\cite{prd96}.

Similarly, taking into account the threshold for the $\pi\Sigma$ sum rule
we can get the mass of the $K^- p$ multiquark state.
The masses are 1.387 GeV and 1.412 GeV for $s_0$ = 3.012 GeV$^2$
and $s_0$ = 3.112 GeV$^2$, respectively. They reproduce
the $\Lambda$ (1405) mass.
On the other hand, the mass of the $\pi^-\Sigma^+$ multiquark state
at the threshold $s_0$ = 3.112 GeV$^2$ is 1.426 GeV.

\section{Discussion}{\label{discuss}}

Let us discuss uncertainties in our sum rules.
First, most uncertainties come from neglecting the contribution of
other dimensional operators (e.g., gluon condensates) on the OPE side.
As we said previously,
there is only one power correction term in our sum rules, thus
the results are
very sensitive to the choice of the continuum threshold $s_0$.
Second, one of uncertainties results from assuming
the vacuum saturation hypothesis
to calculate quark condensates of higher dimensions.
In $\Pi_1$ sum rule the dominant operator has the form of
$\langle\bar{q}q\rangle^3$ while $\langle\bar{q}q\rangle^2$
in $\Pi_q$ sum rule.
Thus, the uncertainty contributes to each sum rule in a different manner.
Last, in the previous $\pi\Sigma$ sum rules we only consider 
the $\pi^0\Sigma^0$
channel. It would be necessary to obtain the sum rules for
the full basis, i.e.,
$\pi^+\Sigma^-$ + $\pi^0\Sigma^0$ + $\pi^-\Sigma^+$ multiquark state.
Full quantitative analysis , however, would require
all the above corrections and is beyond
the scope of this paper.

In summary, the mass of the $\Lambda$(1405) is predicted using
the $\pi^0\Sigma^0$
multiquark interpolating field.
The predicted mass from $\Pi_1$ sum rule
(the chiral-odd sum rule) is about 1.419 GeV.
The mass sum rules for the
$K^+ p$ and the $\pi^+\Sigma^+$ multiquark state are also
presented, and compared to those
for the $\pi^0\Sigma^0$ multiquark state.
It is necessary
to investigate the problem further both theoretically and
experimentally to determine the structure of the $\Lambda$ (1405).
 On the other hand,
it would be interesting  to calculate the mass of the $\Lambda$(1405)
with the baryon and vector meson (e.g., $\Sigma\rho$)
 interpolating field which was proposed
in \cite{strottman79,liu84}.

\acknowledgements

The author thanks Prof. A.W. Thomas, Dr. A.G. Williams,
and Dr. D.B. Leinweber for
useful discussions and comments.
The author wishes to acknowledge the financial support of
the Korea Research Foundation (KRF) made in the program year 1997.
This work is supported in part by
Centre for the Subatomic Structure of Matter (CSSM)
at University of Adelaide.


\begin{table}[t]
\caption{$\pi^0\Sigma^0$ multiquark state
($\langle\bar{s}s\rangle$ = 0.8~$\langle\bar{q}q\rangle$,
$m_s$ = 0.150 GeV) }
\label{table_qc}
\begin{center}
\begin{tabular}{c c c c c}
   & quark condensate (GeV$^3$)
   &	 $s_0$ (GeV$^2$)  & $M^2$ (GeV$^2$)   & $m^\prime$(GeV) \\
\hline
$\Pi_q (q^2)$ &  --(0.210)$^3$ &  3.005     & 0.82 -- 1.68    &  1.303 \\
	      &  --(0.230)$^3$ &  3.122     & 0.98 -- 1.80    &  1.296 \\
	      &  --(0.250)$^3$ &  3.268     & 1.18 -- 1.90    &  1.289 \\
\hline
$\Pi_1 (q^2)$ &  --(0.210)$^3$ &  3.015     & 0.98 -- 1.80    &  1.333 \\
              &  --(0.230)$^3$ &  3.012	    & 1.18 -- 1.90    &  1.331 \\
	      &  --(0.250)$^3$ &  3.008     & 1.40 -- 2.00    &  1.330 \\
\end{tabular}
\end{center}
\end{table}
\begin{table}[t]
\caption{$\pi^0\Sigma^0$ multiquark state
($\langle\bar{q}q\rangle$ = --(0.230 GeV)$^3$,
$\langle\bar{s}s\rangle$ = 0.8~$\langle\bar{q}q\rangle$)}
\label{table_sqm}
\begin{center}
\begin{tabular}{c c c c c}
   & s-quark mass (GeV)
   &	 $s_0$ (GeV$^2$)  & $M^2$ (GeV$^2$)   & $m^\prime$(GeV) \\
\hline
$\Pi_q (q^2)$
       &  0.120    &  3.143	& 1.00 -- 1.80	  &  1.299 \\
       &  0.180    &  3.094	& 0.96 -- 1.80	  &  1.291 \\
\hline
$\Pi_1 (q^2)$
       &  0.120    &  3.030	& 1.20 -- 1.90	  &  1.331 \\
       &  0.180    &  2.996	& 1.16 -- 1.90	  &  1.331 \\
\end{tabular}
\end{center}
\end{table}
\begin{table}[t]
\caption{$\pi^0\Sigma^0$ multiquark state
($\langle\bar{q}q\rangle$ = --(0.230 GeV)$^3$, $m_s$ = 0.150 GeV)}
\label{table_sc}
\begin{center}
\begin{tabular}{c c c c c}
   & s-quark condensate
   &	 $s_0$ (GeV$^2$)  & $M^2$ (GeV$^2$)   & $m^\prime$(GeV) \\
\hline
$\Pi_q (q^2)$ &  0.6~$\langle\bar{q}q\rangle$
	      &  3.146	   & 1.00 -- 1.80    &	1.299 \\
	      &  1.0~$\langle\bar{q}q\rangle$
	      &  3.095	   & 0.96 -- 1.80    &	1.293 \\
\hline
$\Pi_1(q^2)$ &	0.6~$\langle\bar{q}q\rangle$
	      &  2.984	   & 1.16 -- 1.90    &	1.331 \\
	      &  1.0~$\langle\bar{q}q\rangle$
	      &  3.030	   & 1.20 -- 1.90    &	1.331  \\
\end{tabular}
\end{center}
\end{table}
\begin{table}[t]
\caption{Mass of the $\pi^0\Sigma^0$ multiquark state
($\langle\bar{s}s\rangle$ = 0.8~$\langle\bar{q}q\rangle$,
$m_s$ = 0.150 GeV) }
\label{mass_qc}
\begin{center}
\begin{tabular}{c c c c}
   & quark condensate (GeV$^3$) &  $s_0$ (GeV$^2$)  &  $m$(GeV) \\
\hline
$\Pi_1 (q^2)$ &  --(0.210)$^3$ &  3.015   &  1.434 \\
	      &  --(0.230)$^3$ &  3.012   &  1.419 \\
	      &  --(0.250)$^3$ &  3.008   &  1.404 \\
\end{tabular}
\end{center}
\end{table}
\begin{table}[t]
\caption{Mass of the $\pi^0\Sigma^0$ multiquark state
($\langle\bar{q}q\rangle$ = --(0.230 GeV)$^3$,
$\langle\bar{s}s\rangle$ = 0.8~$\langle\bar{q}q\rangle$)}
\label{mass_sqm}
\begin{center}
\begin{tabular}{c c c c}
   & s-quark mass (GeV)  &  $s_0$ (GeV$^2$)  & $m$(GeV) \\
\hline
$\Pi_1 (q^2)$
       &  0.120    &  3.030	&  1.419 \\
       &  0.180    &  2.996	&  1.419 \\
\end{tabular}
\end{center}
\end{table}
\begin{table}[t]
\caption{Mass of the $\pi^0\Sigma^0$ multiquark state
($\langle\bar{q}q\rangle$ = --(0.230 GeV)$^3$, $m_s$ = 0.150 GeV)}
\label{mass_sc}
\begin{center}
\begin{tabular}{c c c c}
   & s-quark condensate  &  $s_0$ (GeV$^2$)  & $m$(GeV) \\
\hline
$\Pi_1(q^2)$ &	0.6~$\langle\bar{q}q\rangle$
	      &  2.984	   & 1.417 \\
	      &  1.0~$\langle\bar{q}q\rangle$
	      &  3.030	   & 1.419 \\
\end{tabular}
\end{center}
\end{table}
\begin{table}[t]
\caption{$K^+ p $ multiquark state
($\langle\bar{q}q\rangle$ = --(0.230 GeV)$^3$,
$\langle\bar{s}s\rangle$ = 0.8~$\langle\bar{q}q\rangle$,
$m_s$ = 0.150 GeV)}
\label{table_kp}
\begin{center}
\begin{tabular}{c c c c}
   &	  $s_0$ (GeV$^2$)  & $M^2$ (GeV$^2$)   & $m^\prime$(GeV) \\
\hline
$\Pi_q (q^2)$ &  3.646	   & 1.06 -- 2.00    &	1.372 \\
\hline
$\Pi_1 (q^2)$ &  3.852	   & 0.94 -- 2.22    &	1.440\\
\end{tabular}
\end{center}
\end{table}
\begin{table}[t]
\caption{$\pi^+\Sigma^+$ multiquark state
($\langle\bar{q}q\rangle$ = --(0.230 GeV)$^3$,
$\langle\bar{s}s\rangle$ = 0.8~$\langle\bar{q}q\rangle$,
$m_s$ = 0.150 GeV)}
\label{table_ps}
\begin{center}
\begin{tabular}{c c c c}
   &	  $s_0$ (GeV$^2$)  & $M^2$ (GeV$^2$)   & $m^\prime$(GeV) \\
\hline
$\Pi_q (q^2)$ &  3.126	   & 0.98 -- 1.80    &	1.239 \\
\hline
$\Pi_1 (q^2)$ &  3.112	   & 1.20 -- 1.88    &	1.330 \\
\end{tabular}
\end{center}
\end{table}

\begin{figure}
\caption{Mass of the $\pi^0\Sigma^0$ multiquark state at
the continuum threshold
$s_0$ = 2.789 GeV$^2$.	The solid line is the predicted mass from
$\Pi_q$ sum rule, and the dotted line is that from $\Pi_1$ sum rule.}
\label{fig_ps0s}
\end{figure}

\begin{figure}
\caption{Dependence of the mass of the $\pi^0\Sigma^0$ multiquark state
from $\Pi_1$ sum rule on
~(a) strange quark mass ~(b) strange quark condensate
~(c) quark condensate .}
\label{fig_ps0v}
\end{figure}

\begin{figure}
\caption{Diagrams. Solid lines are the quark propagators.
~(a) 2 loop-type ~(b) 1 loop-type .}
\label{fig_loop}
\end{figure}

\begin{figure}
\caption{Mass of the $\pi^0\Sigma^0$ multiquark state in the valid
Borel region.
$m$ is the mass with 2 loop-type
diagrams, and $m^\prime$ is the mass with all diagrams
(1 loop-type + 2 loop-type). ~(a) $\Pi_q$ sum rule
~(b) $\Pi_1$ sum rule.}
\label{fig_ps0}
\end{figure}

\begin{figure}
\caption{Coupling strength $\lambda^2$ from $\Pi_q$ and $\Pi_1$ sum rule.}
\label{fig_coup}
\end{figure}

\begin{figure}
\caption{Mass of the $K^+ p$ multiquark state in the valid Borel
region. The same as Fig. 4.}
\label{fig_kp}
\end{figure}

\begin{figure}
\caption{Mass of the $\pi^+\Sigma^+$ multiquark state  in the valid Borel
region. The same as Fig. 4.}
\label{fig_ps}
\end{figure}


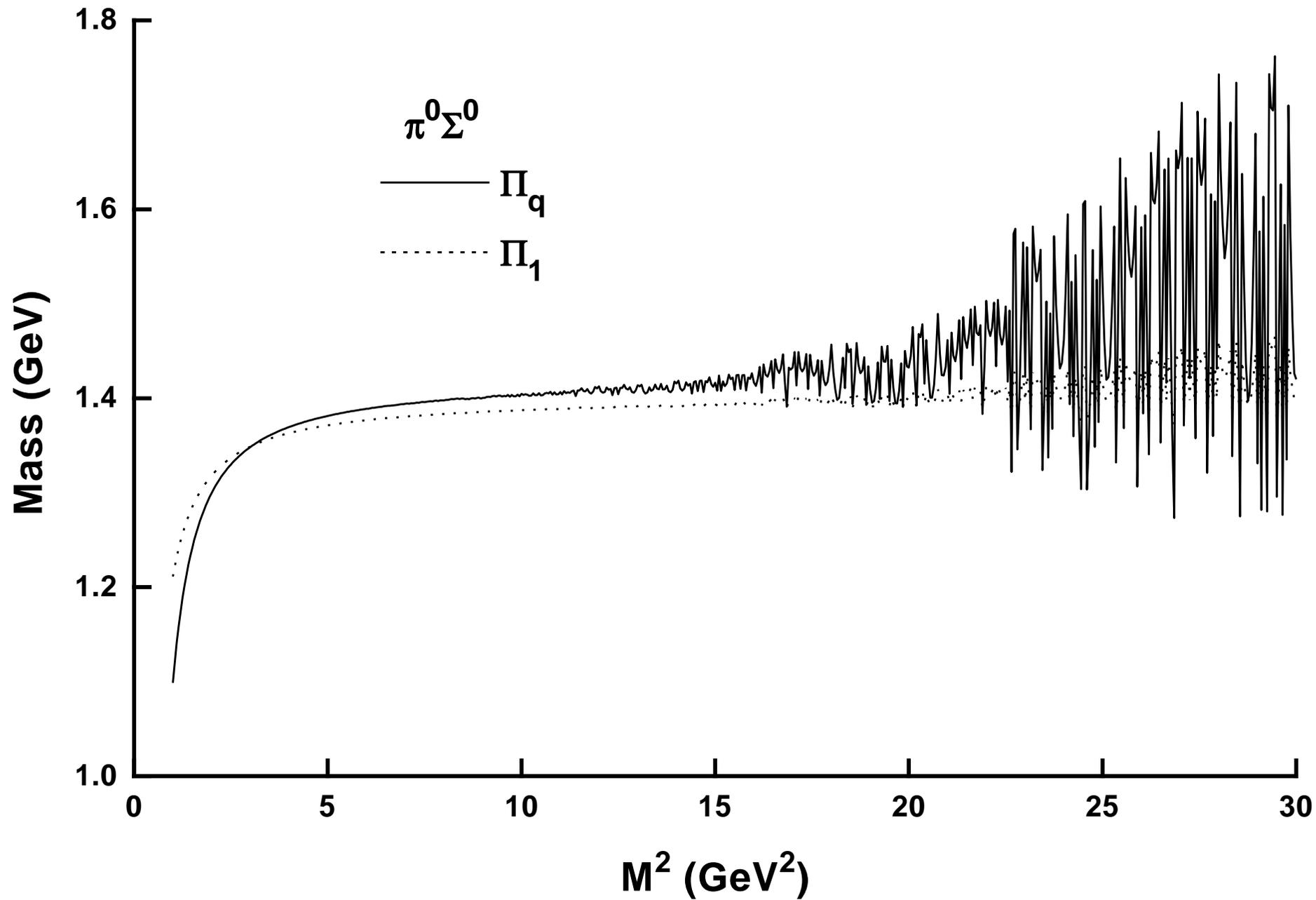
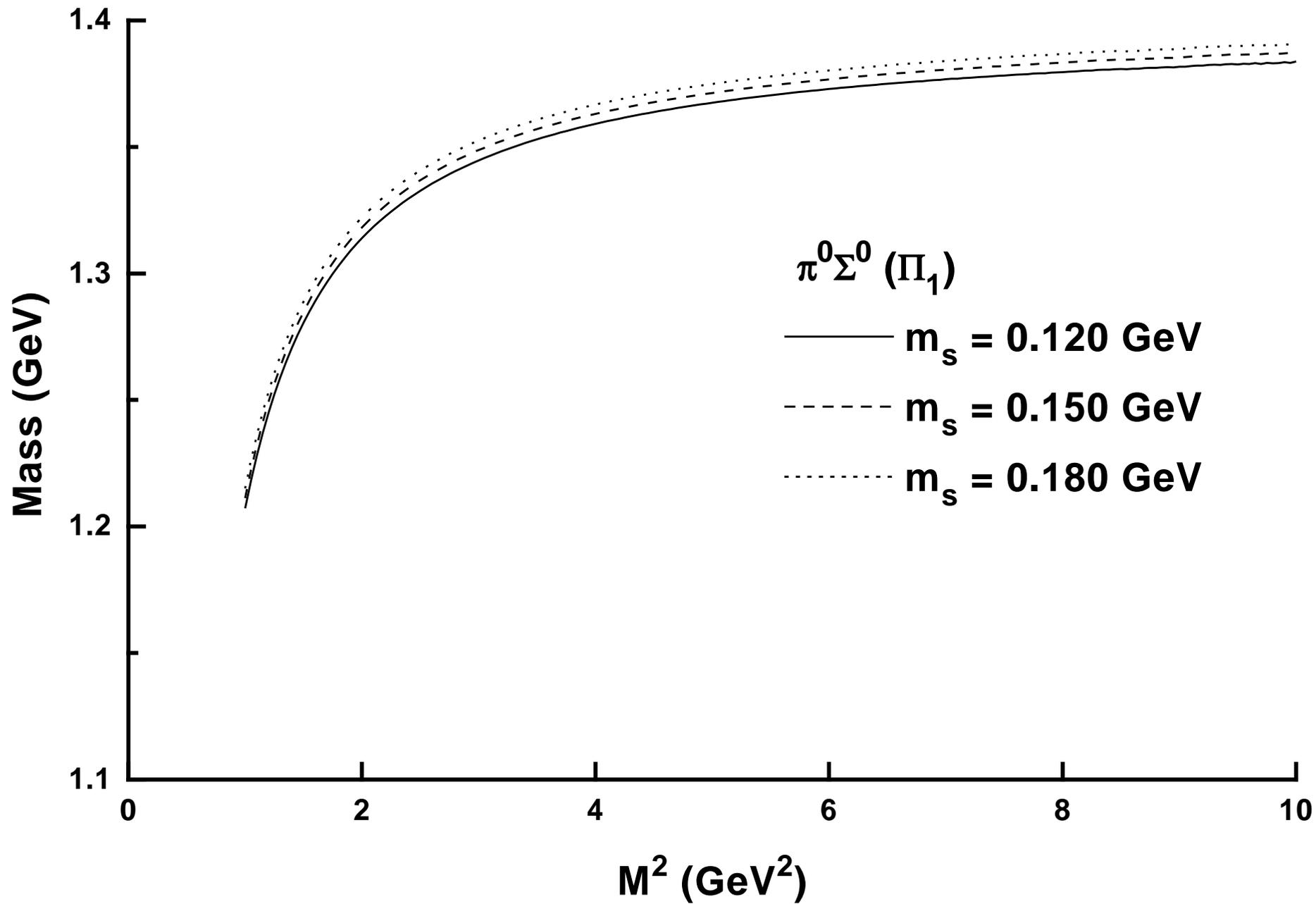
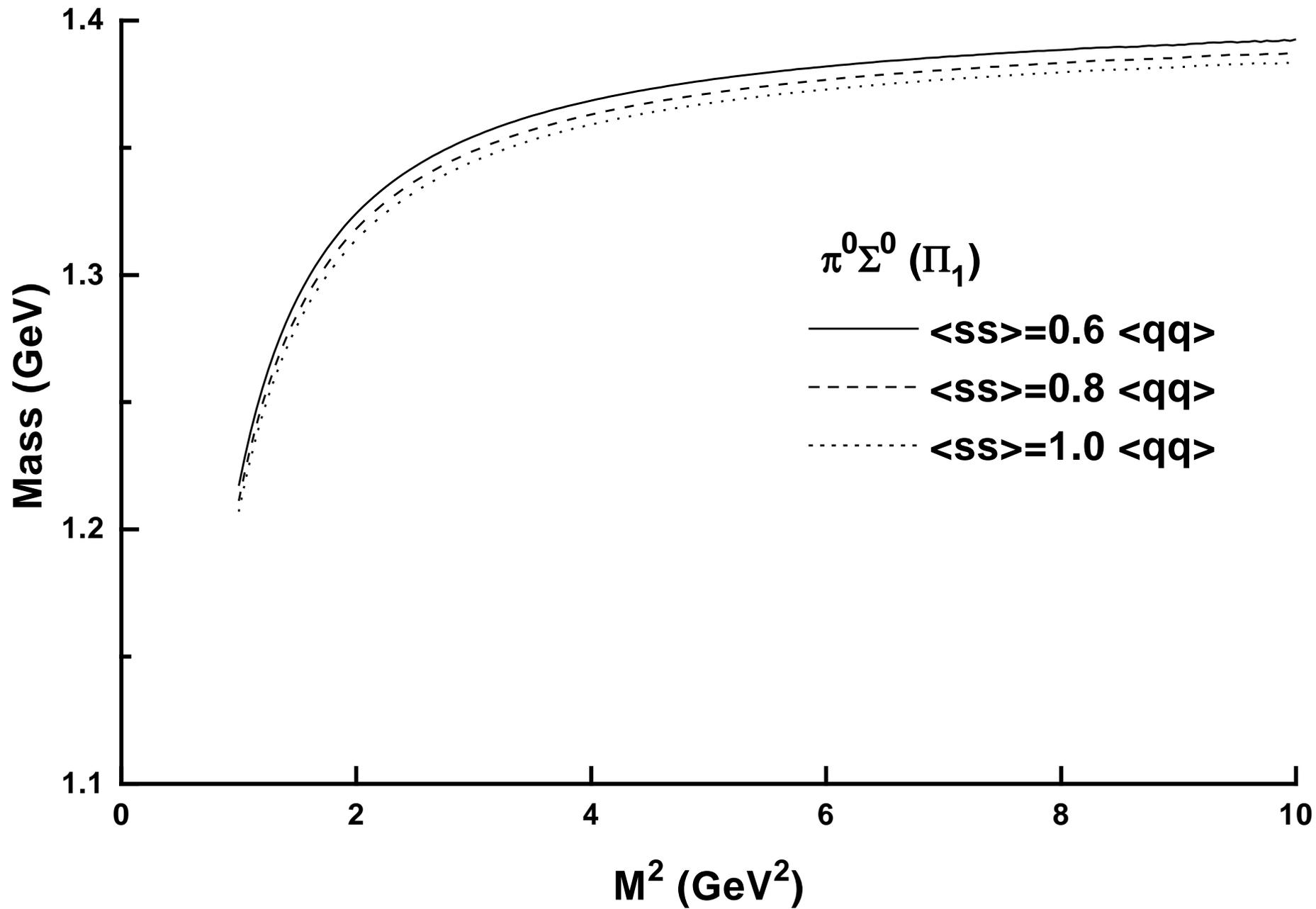
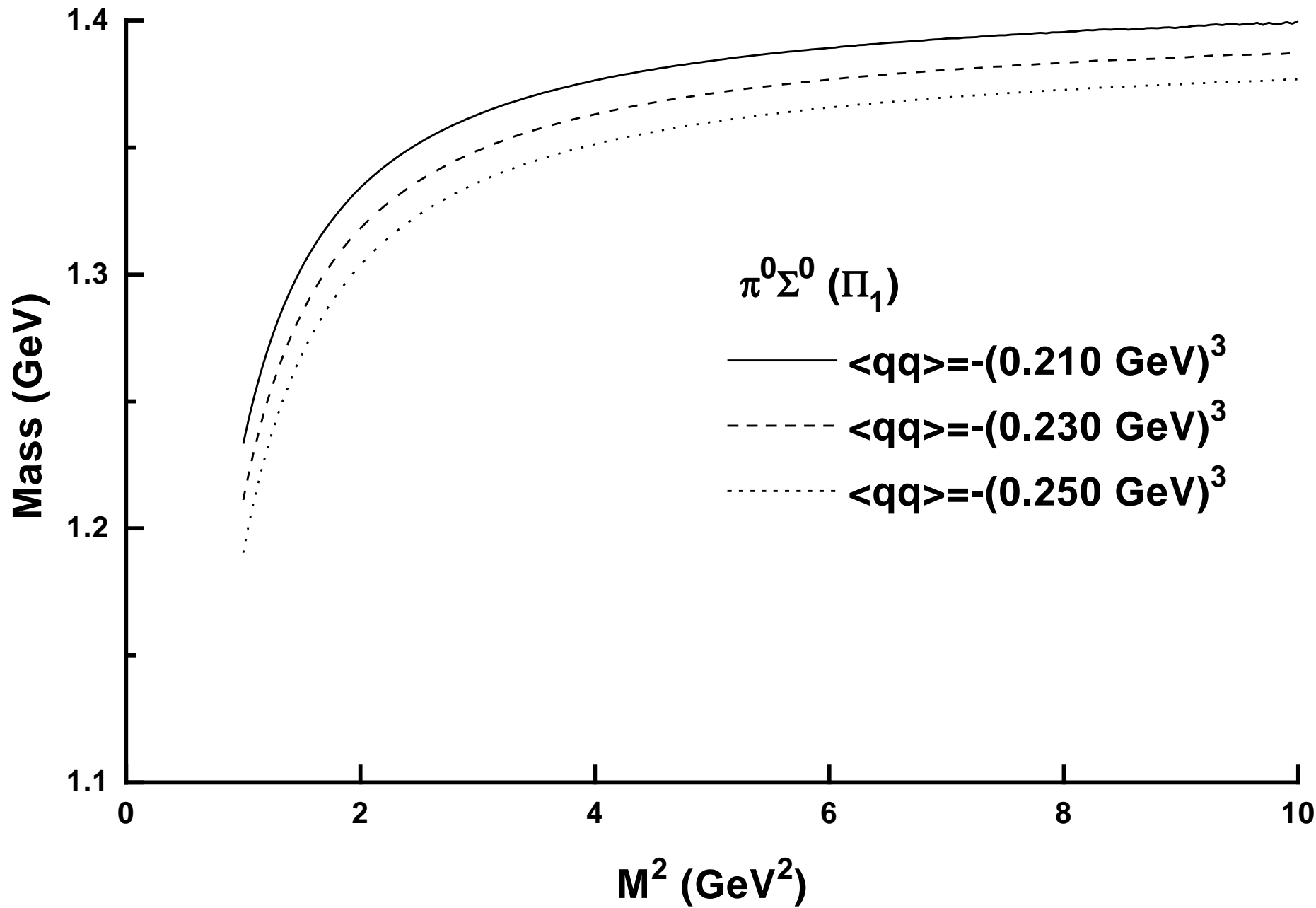
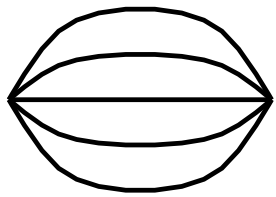
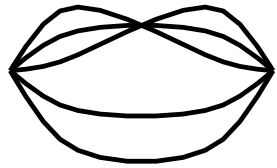
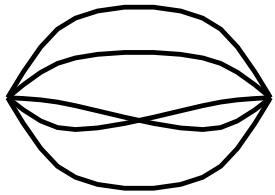
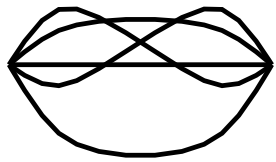
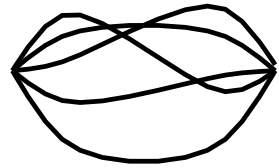
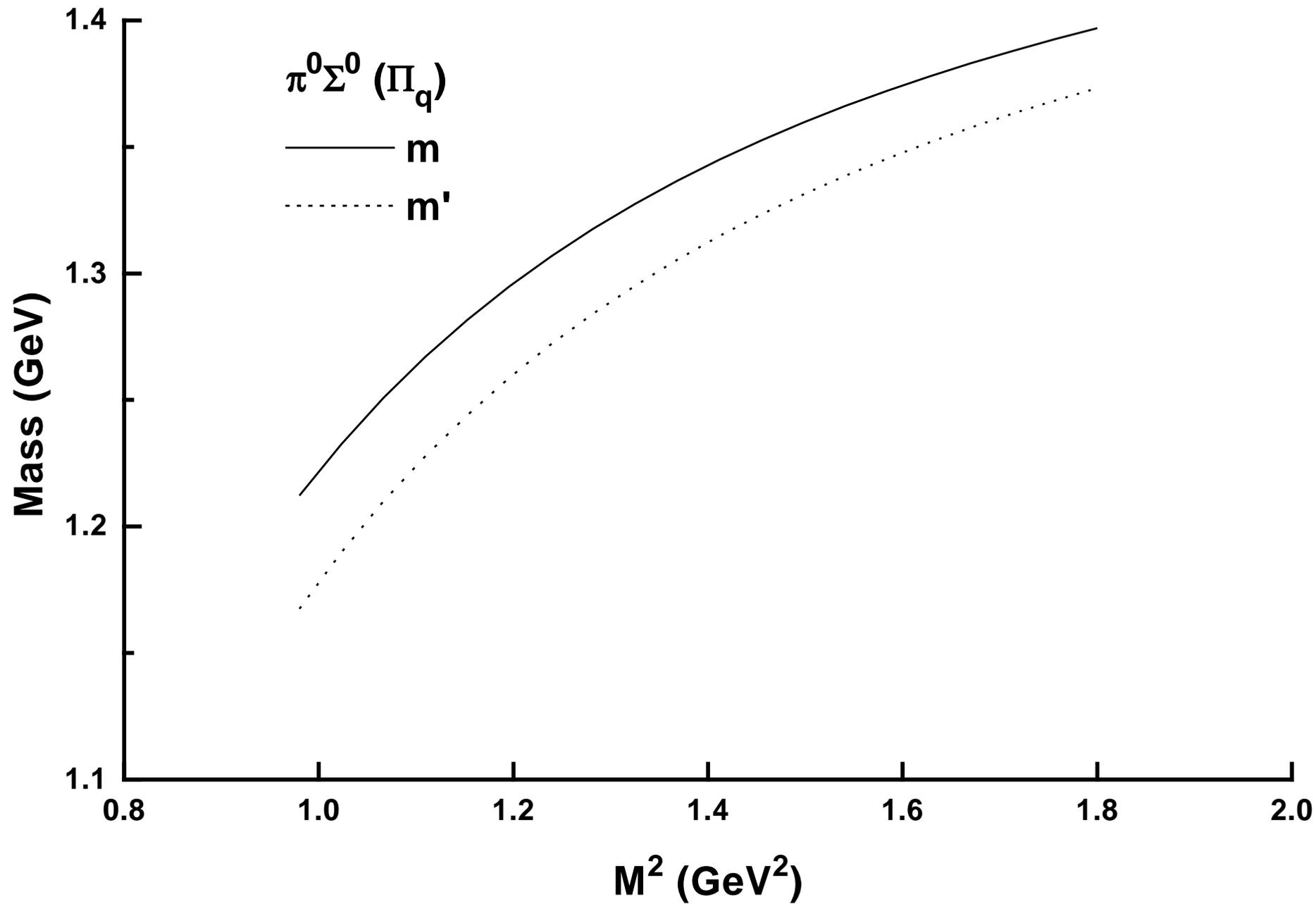
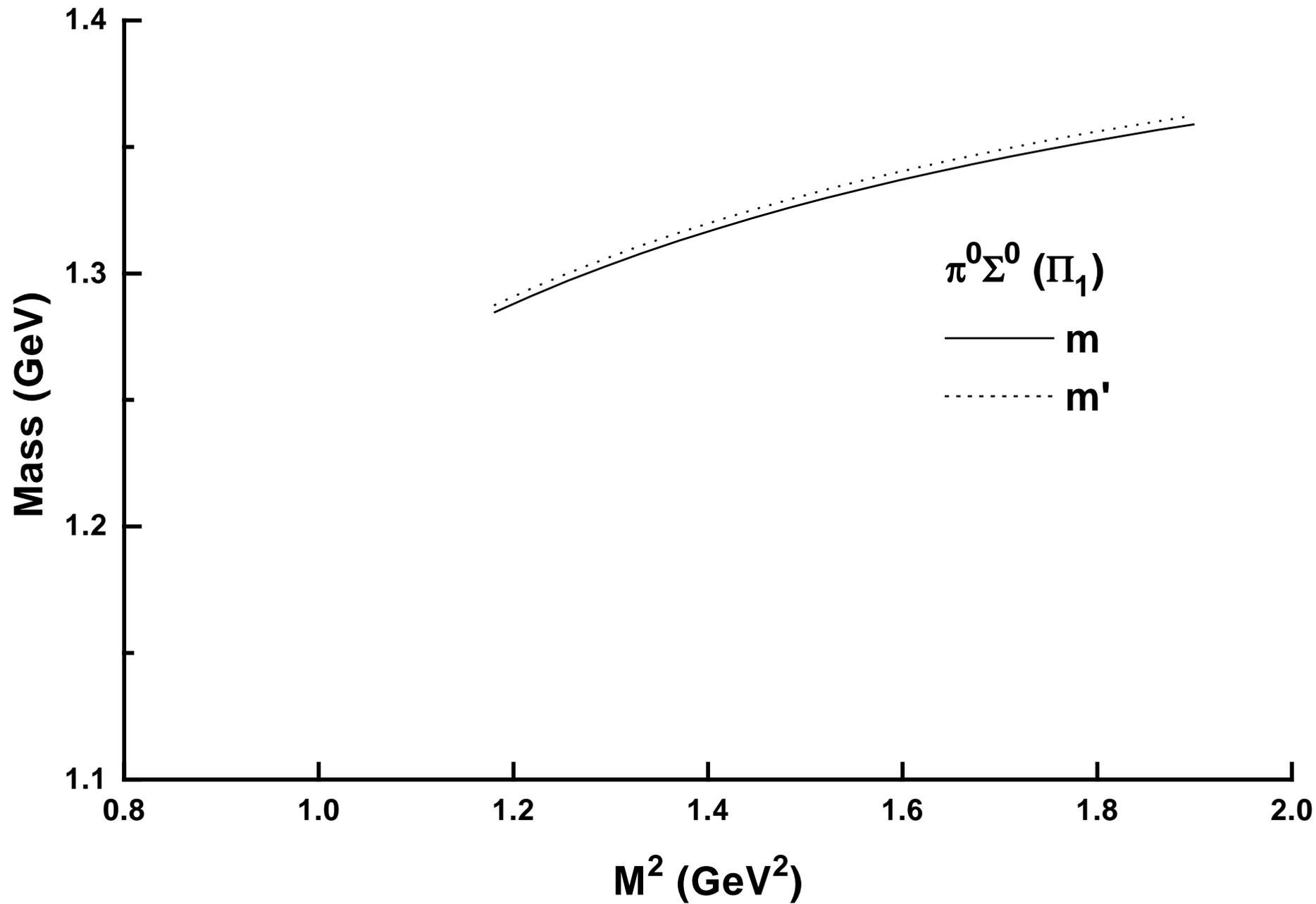
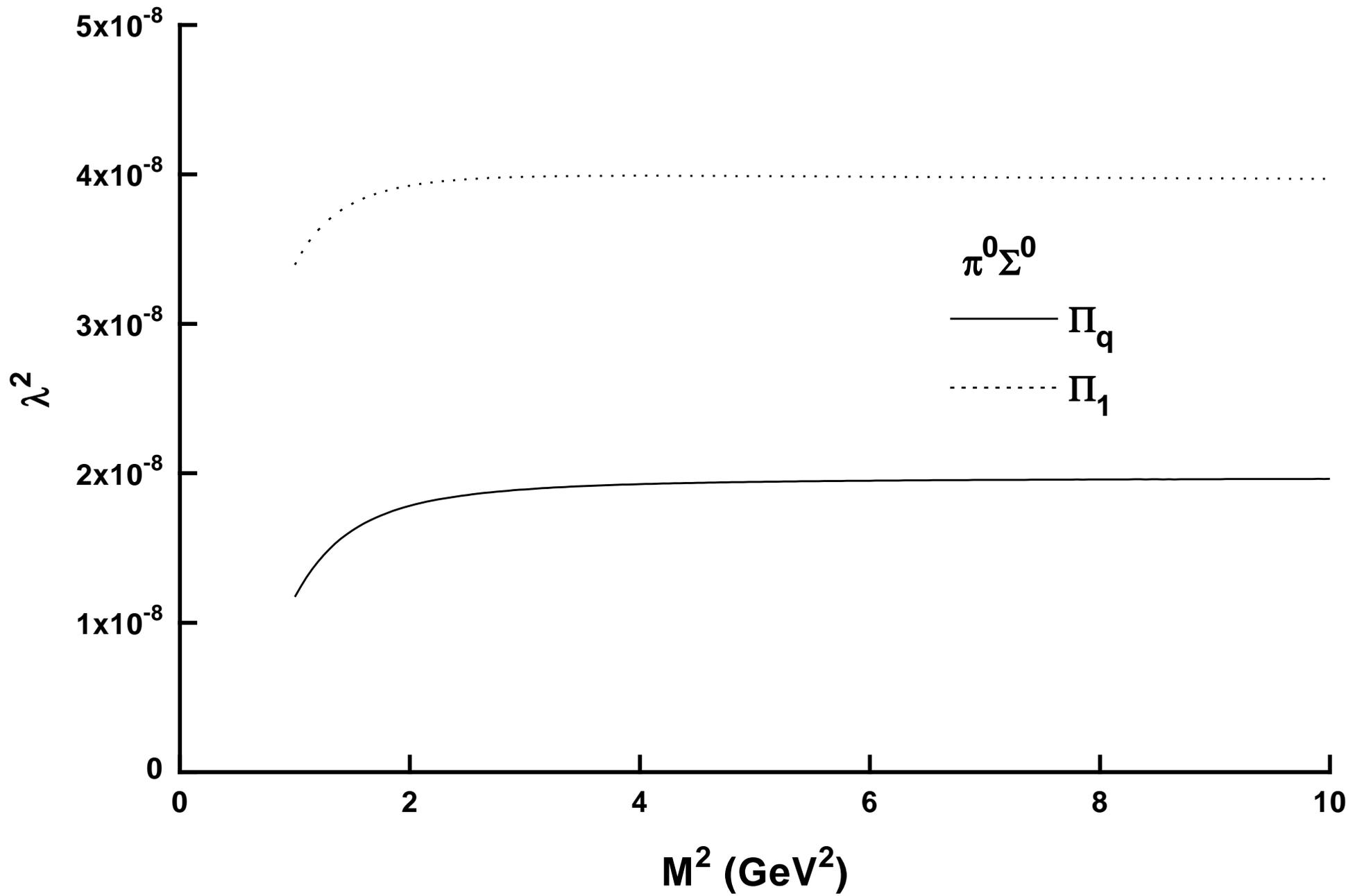
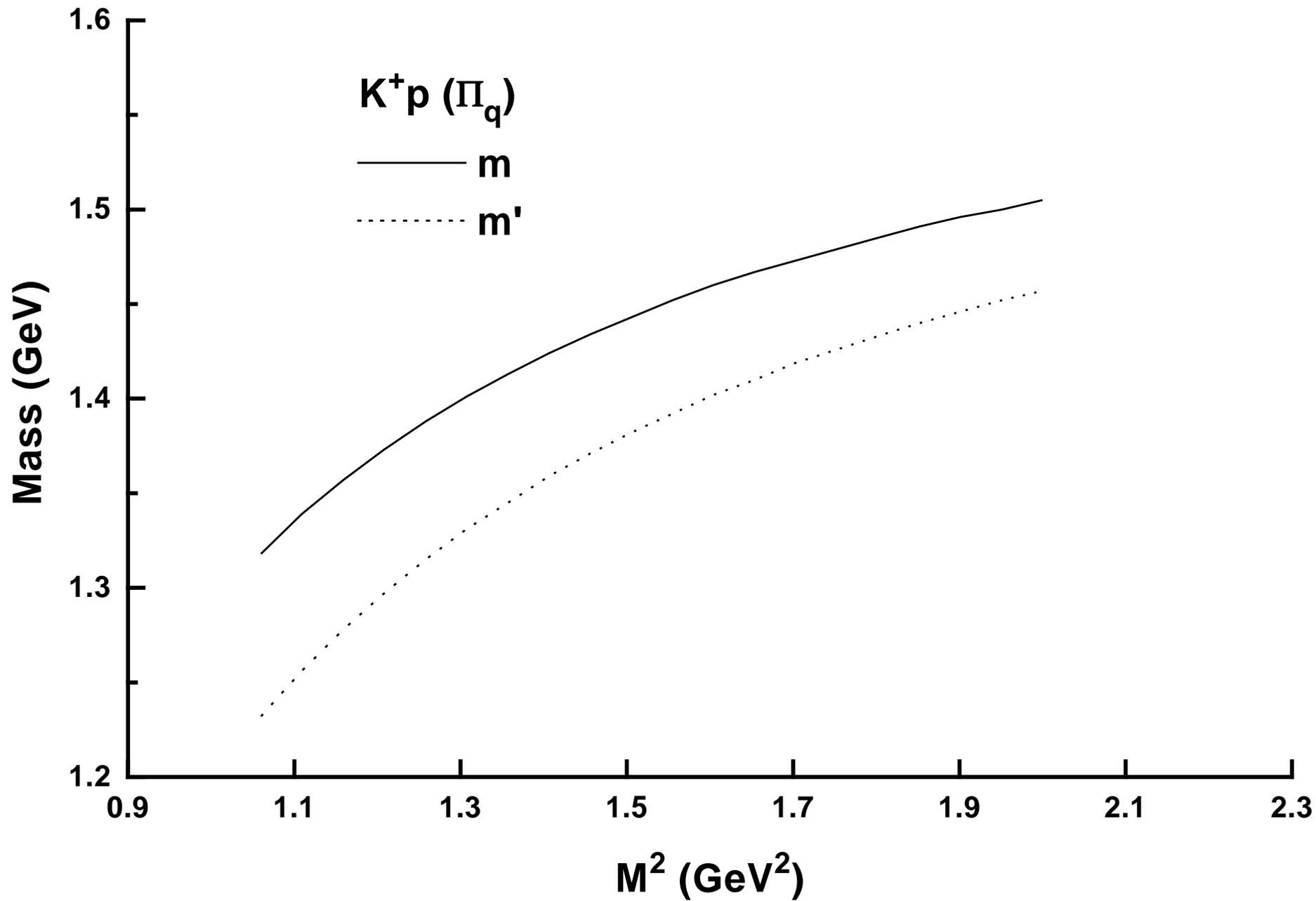
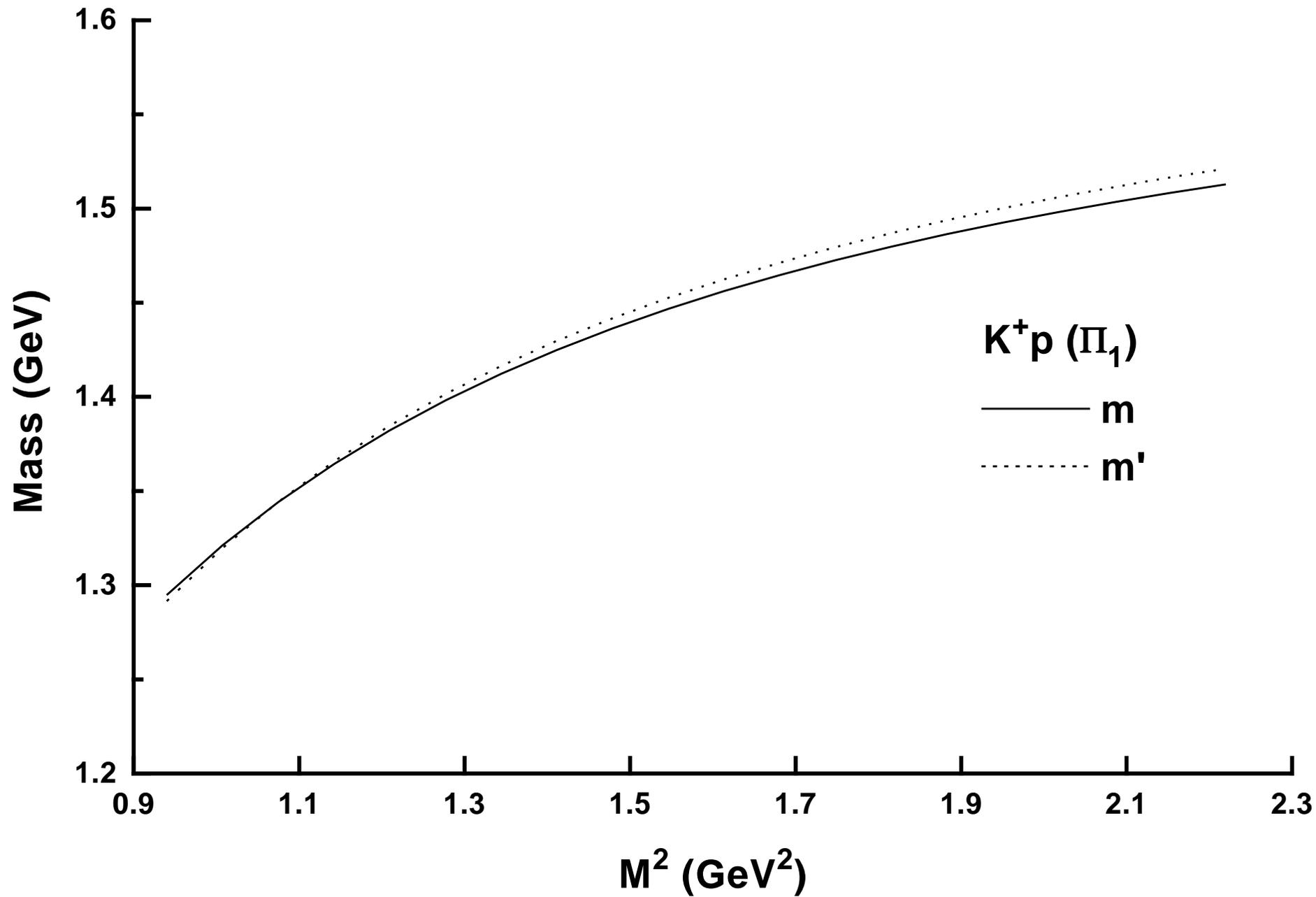
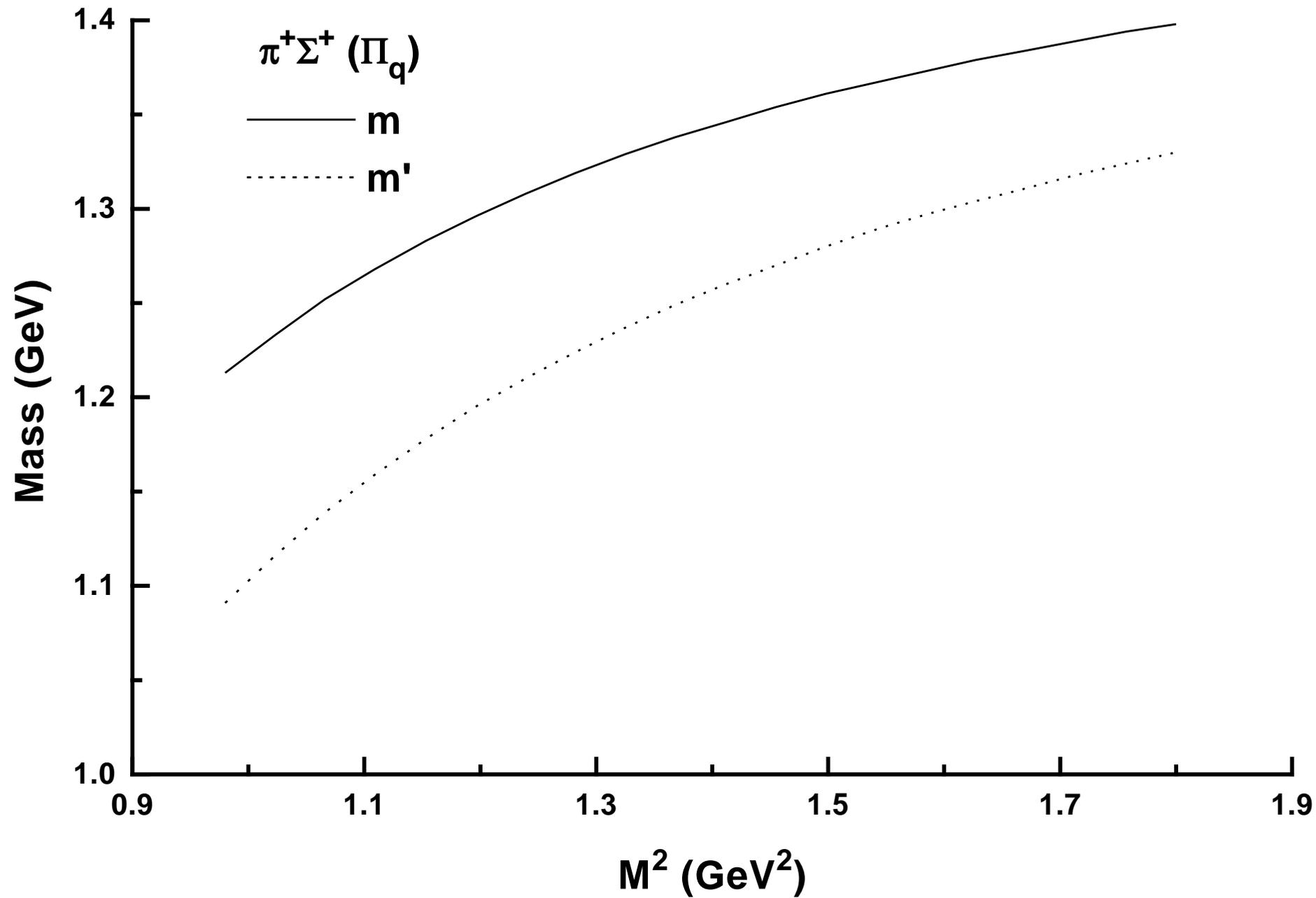
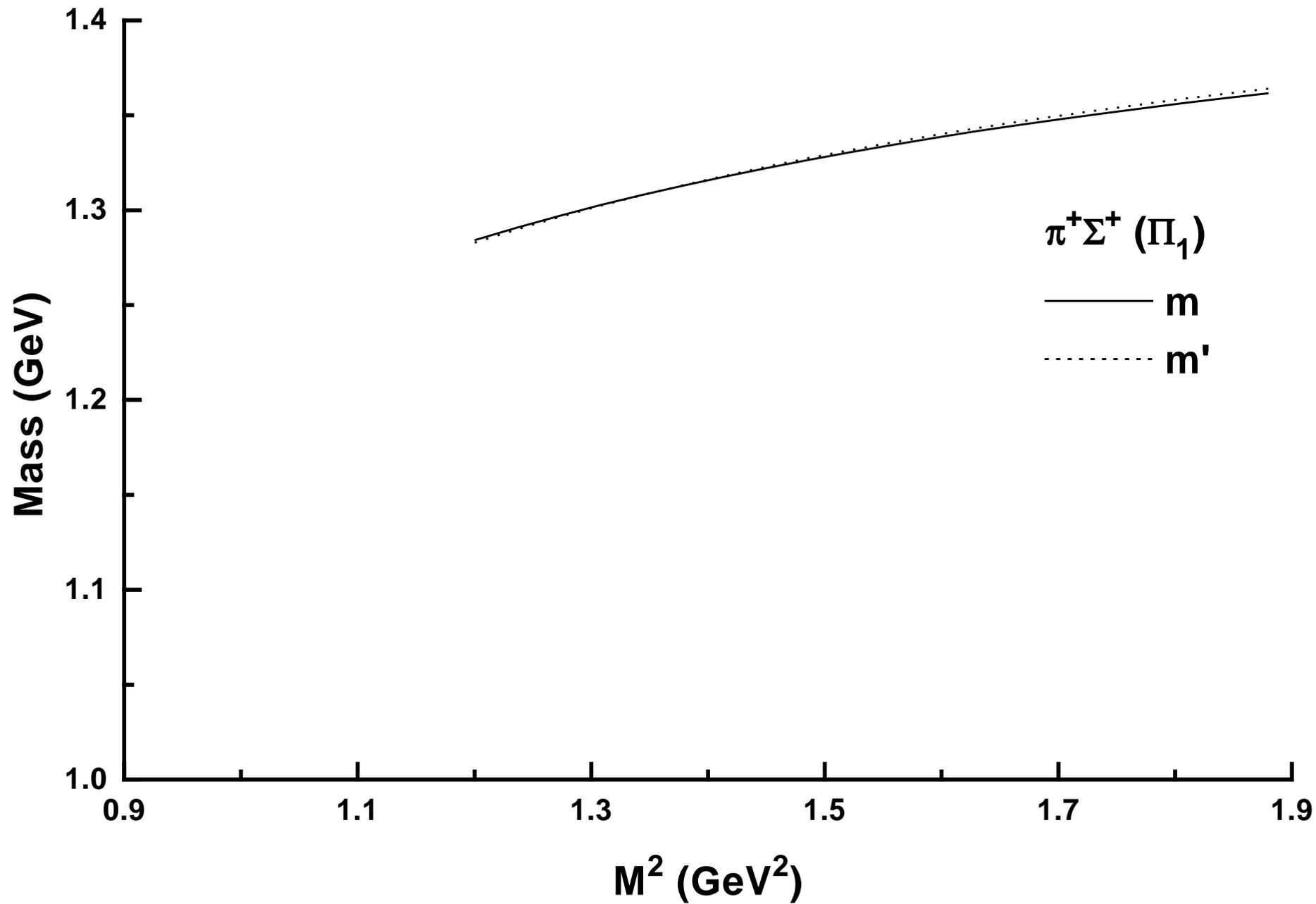
\end{document}